\documentclass[11pt,a4]{article}
\usepackage{amsmath}
\usepackage{amsfonts}
\usepackage{amssymb}
\usepackage{epsfig}
\usepackage{times}
\def\beq{\begin{equation}}
\def\eeq{\end{equation}}
\def\beeq{\begin{eqnarray}}
\def\eeeq{\end{eqnarray}}
\def\cO#1{{\cal{O}}\left(#1\right)}

\def\tw{\textwidth}

\begin{document}
\begin{center}
\begin{large}
\textbf{\Large Correlations between two particles in jets}
\end{large}
\vskip 0.2cm
Redamy Perez Ramos 
\vskip 0.3cm 
Laboratoire de Physique Th\'eorique et Hautes Energies,
Universit\'es Paris 6 et 7,
\newline 4 place Jussieu F-75251 PARIS
C\'edex 05 
\vskip 0.3cm
\end{center}

\textbf{Abstract:} 
We study the correlation between two soft particles in QCD jets. We
extend the Fong--Webber analysis to the region away from the hump in
the single inclusive energy spectrum and show that the correlation
function should flatten off and then decrease for large values of
$\ell=\ln\left(1/x\right)$.

%


\vspace {5mm}
\noindent
{\large\bf Introduction}

Perturbative QCD (pQCD)
successfully 
predicted 
inclusive particle spectra in jets (for review see~\cite{KO}
and references therein).
It sufficed to make one step beyond the leading double log
approximation (DLA) and to analyse parton cascades with the
next-to-leading accuracy by taking into account angular ordering in
soft gluon multiplication, quark--gluon transitions, exact DGLAP
parton splitting functions and running coupling effects.  The
corresponding MLLA (modified leading log approximation) ``{\em
  evolution Hamiltonian}\/'' has the accuracy $\gamma_0+\gamma_0^2$,
where $\gamma_0$ is the characteristic parameter of the pQCD expansion
(the DLA multiplicity anomalous dimension) given by
$$
  \gamma_0^2 \>=\> \frac{4N_c\,\alpha_s(k^2_{\perp})}{2\pi} =
  \frac{4N_c}{\beta\, Y}\,; \quad
  \beta=\frac{11N_c}{3} -  \frac{4}{3}T_R\,, \>\> T_R=\frac12 n_f.
$$
Here $Y=\ln (Q/2\Lambda)$, with $Q$ the hardness of the jet production
process ($e^+e^-$ c.m.s.\ annihilation energy) and $\Lambda$ the QCD
scale.  For the LEP-I energy, $Q=91.2$ GeV ($Z^0$ peak) we have
$Y\simeq 5.2$ (using $\Lambda\simeq 0.25$ GeV).

\medskip
\noindent
\begin{minipage}{0.55\tw}
 \epsfig{file=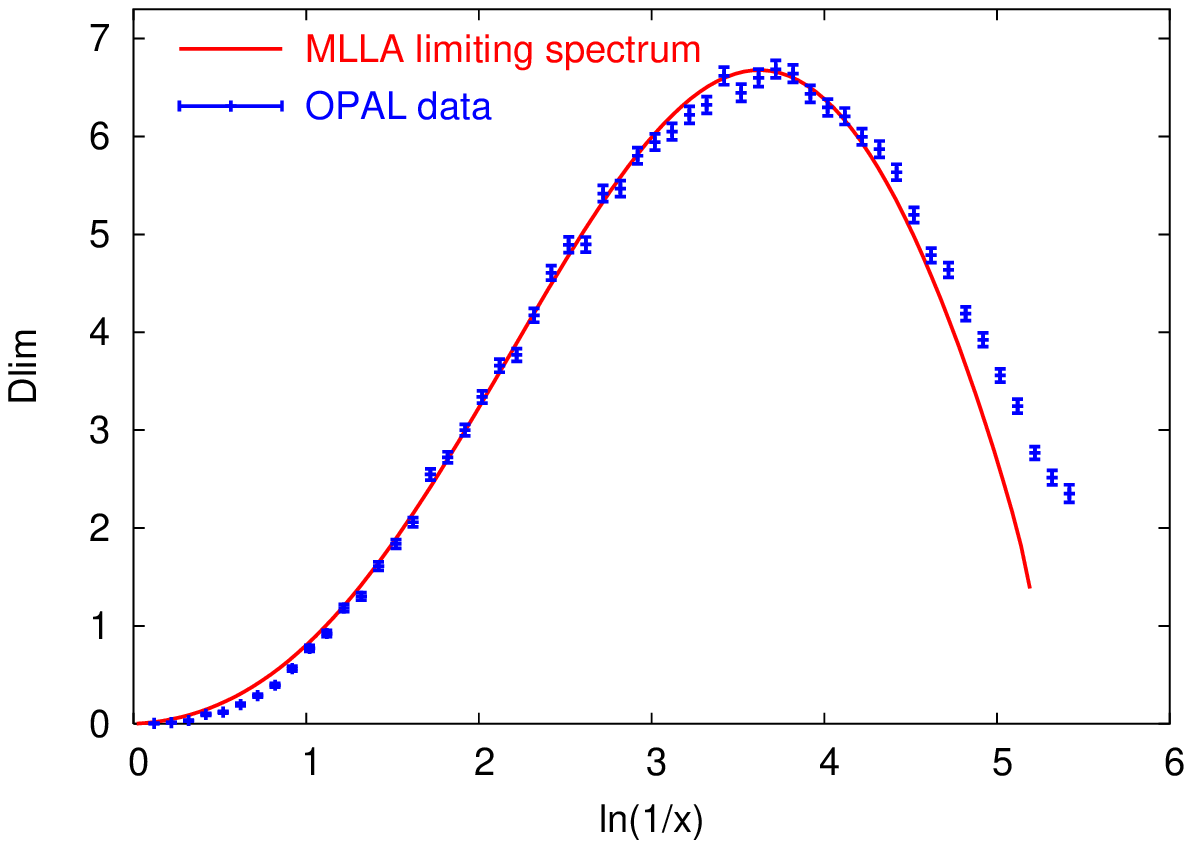,width=\tw,height=0.95\tw} \vskip
  .1cm {Fig.1: Inclusive spectrum for $Q=91.2$ GeV.}
\end{minipage}
\hfill
\begin{minipage}{0.4\tw}
In spite of expansion parameter being numerically large, 
\beq\label{g0val}
\gamma_0\simeq 0.5\,,
\eeq
the {\em shape}\/ of the inclusive energy distribution of soft
hadrons turned out to be well described by the analytic QCD curve ---
the so-called limiting spectrum 
$$ D^{\mbox{\scriptsize lim}} \left(\ell, Y\right) =
\frac{1}{\sigma}\, \frac{d\sigma}{d\ell}\,; \quad\ell =\ln\frac1x\,,
$$
that one derives pushing the minimal transverse momentum in partonic
cascades $Q_0$ down to $\Lambda$.
\end{minipage}
\medskip 

Overall normalization in Fig.~1 (number of hadrons per gluon) is a
non-per\-tur\-ba\-tive parameter that one determines
phenomenologically.  This parameter should cancel in the ratio
\beq\label{corrdef} C_{G,Q}(\ell_1,\ell_2,Y)\>=\>
\frac{D_{G,Q}^{\left(2\right)}\left(\ell_1,\ell_2, Y\right)}
{D_{G,Q}\left(\ell_1, Y\right)\,D_{G,Q} \left(\ell_2, Y\right)} \,,
\eeq
so that one could expect the two-particle correlation function
\eqref{corrdef} to provide a more stringent test of parton dynamics.
However, for a long time, particle correlations are known to be poorly
treated by pQCD.

Here we revisit two-particle correlations and report the results of an
improved pQCD analysis of the problem~\cite{DPR}.


\vspace {5mm}
\noindent
{\large\bf Fong--Webber approximation}

The first (and only) pQCD analysis of two-particle correlations in jets
beyond the DLA was performed by Fong and Webber in 1990. In~\cite{FW}
the next-to-leading $\cO{\gamma_0}$ correction,
$C=1+\sqrt{\alpha_s}+\cdots$, to the normalized two-particle
correlator was calculated.
For a system of two quark jets produced in $e^+e^-$ annihilation,
$$
 R\left(\ell_1, \ell_2, Y\right) \>=\>
 \frac1{2}+\frac1{2}C_Q\left(\ell_1, \ell_2, Y\right),
$$
the Fong--Webber result reads ($n_f=3$)
\beq\label{FWans}
R
\>=\>
1.375-1.125\left(\frac{\ell_1-\ell_2}{Y}\right)^2-\left[1.262 -
  0.877\frac{\left(\ell_1+\ell_2\right)}{Y}\right]\frac1{\sqrt{Y}}.
\eeq 
The first two terms in \eqref{FWans} are of the DLA origin while the
third one constitutes an $\cO{\gamma_0}$ MLLA correction.
This expression was derived in the region $|\ell_1-\ell_2|/Y \ll 1$,
that is when the energies of the registered particles are relatively
close to each other (and to the maximum of the inclusive
distribution, see Fig.1).  In this approximation the correlation function is
quadratic in the difference $\ell_1-\ell_2$ and increases linearly
with the sum, $\ell_1+\ell_2$.



\vspace {5mm}
\noindent
{\large\bf Particle correlations from MLLA evolution equations}

To analyse two-particle correlations we write down the evolution
equations for $D_{G,Q}^{(2)}\left(\ell_1,\ell_2, Y\right)$ that follow
from the general MLLA evolution equations for jet generating
functionals~\cite{EvEq}. Unlike the case of the
inclusive spectrum, equations for
 $D^{(2)}$
 are inhomogeneous due to the presence of the product of one-particle
 distributions $D(\ell_1)D(\ell_2)$.  In the small-$x$ limit, by
 approximating the energy fraction integrals of the parton splitting
 functions as follows,
$$
 \int_0^\ell d\ell' \> P(x')\, F(\ell-\ell') \>\simeq\> \int_0^\ell
 d\ell' \> \left[\,c_1 - c_2\delta(\ell')\,\right]\, F(\ell-\ell')
\,,
$$
it is straightforward to reduce the original integral equations to
a system of two linear (inhomogeneous) second order differential
equations in variables $\ell_1$ and $y_2$, where $y\equiv
\ln(k_\perp/\Lambda)$ and we keep the ratio of particle energies
fixed, $\ell_1-\ell_2=$const.

We then substitute the product $C\cdot D(\ell_1)D(\ell_2)$ for
$D^{(2)}$ and solve the equations iteratively, using the fact that the
normalized correlator \eqref{corrdef} is a slowly changing function as
compared with the distributions themselves. The range of applicability
of the solution so obtained is not restricted to the vicinity of the
hump as in~\cite{FW}. Actually, the new solution can be trusted for
arbitrary values of $\ell_i$, as long as $\ell_1,\ell_2>2$ (to respect
the adopted soft approximation, $x\ll1$).

For example, the answer for the two-particle correlation inside a
gluon jet reads 
\beq\label{eq:Cglufull} 
C_G\left(\ell_1, \ell_2, Y\right)-1 = \frac{1-b\left(\psi_{1,\ell}
    +\psi_{2,\ell}-[\beta\gamma_0^2]\right)
  -\delta_1-[a\chi_{\ell}+\delta_2]} {1+ \Delta
  + \delta_1 + [a(\chi_{\ell}+\beta\gamma_0^2)+\delta_2]}.  
\eeq 
The function $\Delta
$ is given by
\beq\label{Deltadef}
 \Delta \>=\>
 \gamma_0^{-2}\left(\psi_{1,\ell}\psi_{2,y}+\psi_{1,y}\psi_{2,\ell}\right),
\eeq
where $\psi_i=\ln D(\ell_i,Y)$, and the subscripts $\ell$ and $y$ mark
its respective derivatives.  Since logarithmic derivatives of the
inclusive spectrum are, typically, of the order of
$\psi_{\ell}\sim\psi_y = \cO{\gamma_0}$, we have $\Delta = \cO{1}$.
This term is already present in the DLA, see~\cite{DLA}, and is
responsible for the fall-off of the correlation with increase of
$\eta=|\ell_1-\ell_2|$.

The term $b\left(\psi_{1,\ell}+\psi_{2,\ell}\right)$, 
where $b=\frac1{4N_c}\left[\frac{11}3 N_c -
  \frac{4}{3}T_R \left(1-\frac{2C_F}{N_c}\right)^2 \right]$ in the
numerator of \eqref{eq:Cglufull} is the next-to-leading (MLLA)
correction $\cO{\gamma_0}$.
The term  $\delta_1$  is
given by
\beq\label{delta1} 
 \delta_1 =\gamma_0^{-2}\left[\,\chi_{\ell}(\psi_{1,y} \!+\! \psi_{2,y}) +
 \chi_{y}(\psi_{1,\ell}\!+\!\psi_{2,\ell})\,\right], \qquad
 \chi\equiv \ln[C_G].  
\eeq 
It is $\cO{\gamma_0}$ and constitutes a MLLA correction as well, since
$\chi_{\ell}\sim\chi_y=\cO{\gamma_0^2}$.
%

The correction term $\delta_2$ is given by 
\beq\label{delta2}
\delta_2=\gamma_0^{-2}\left(\chi_{\ell}\chi_y+\chi_{\ell y}\right),
\qquad \chi_{\ell y}\sim \chi_\ell\chi_y = \cO{\gamma_0^4}.  
\eeq 
It is $\cO{\gamma_0^2}$ and constitutes a NMLLA correction, as well as
the term $a\chi_{\ell}$.  Thus, all terms in the square brackets in
\eqref{eq:Cglufull} are of the order of $\gamma_0^2$. Being formally
negligible within the MLLA accuracy, we nevertheless keep them in the
answer. By so doing we follow the logic that was used in the
derivation of the single inclusive MLLA spectra namely, solving {\em
  exactly}\/ the {\em approximate}\/ (MLLA) evolution equations.

The correlator $C_G$ itself enters on the r.h.s.\ of
\eqref{eq:Cglufull} via \eqref{delta1} and \eqref{delta2}. Substituting
\beq\nonumber 
\chi=\ln[C_G] \simeq \ln\left\{1+\frac{1-b\left(\psi_{1,\ell}
      +\psi_{2,\ell}-[\beta\gamma_0^2]\right) ]} {1+\Delta
    + [a\beta\gamma_0^2]}\right\} 
\eeq 
into the expressions for $\delta_1$ and $\delta_2$ provides then an
{\em iterative solution}\/.  The sum of the correction terms
$\delta_1+\delta_2+a\chi_{\ell}$ is very small, $\cO{10^{-2}}$, near
the hump ($\ell_1\simeq\ell_2$) and increases away from it. 

The correlators for the quark jet and for $e^+e^-$ annihilation (two
quark jets) can be obtained from that of the gluon jet.

Below the results are presented for four different bands of the
$\left(\ell_1,\ell_2\right)$ plane, in comparison with the
Fong--Webber approximation and the OPAL data~\cite{OPAL}.  In the left
column we show $R-1$ as a function of the sum $\ell_1+\ell_2$ for three
values of the difference, and on the right column, vice
versa, as a function of the difference for fixed sum.

\begin{center}
\epsfig{file=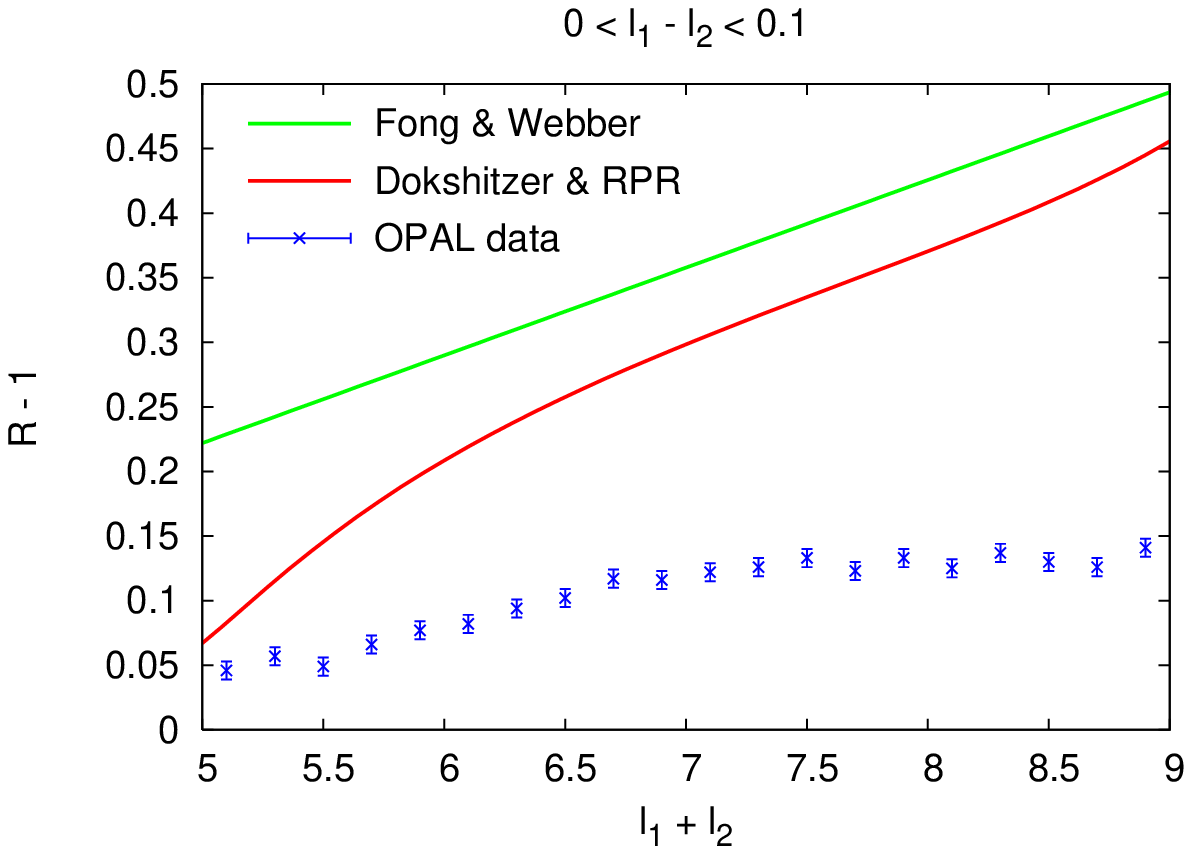, height=4truecm,width=0.45\tw}
\hfill
\epsfig{file=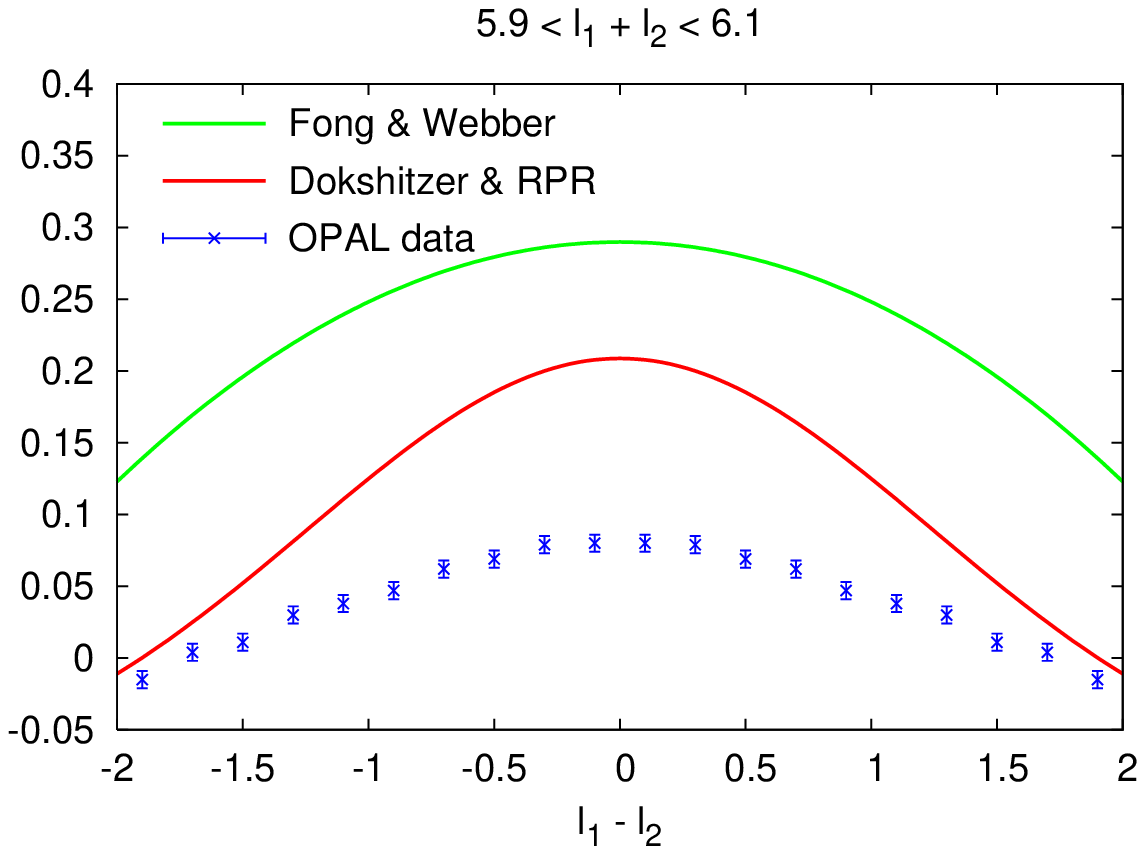, height=4truecm,width=0.45\tw}
\end{center}

\begin{center}
\epsfig{file=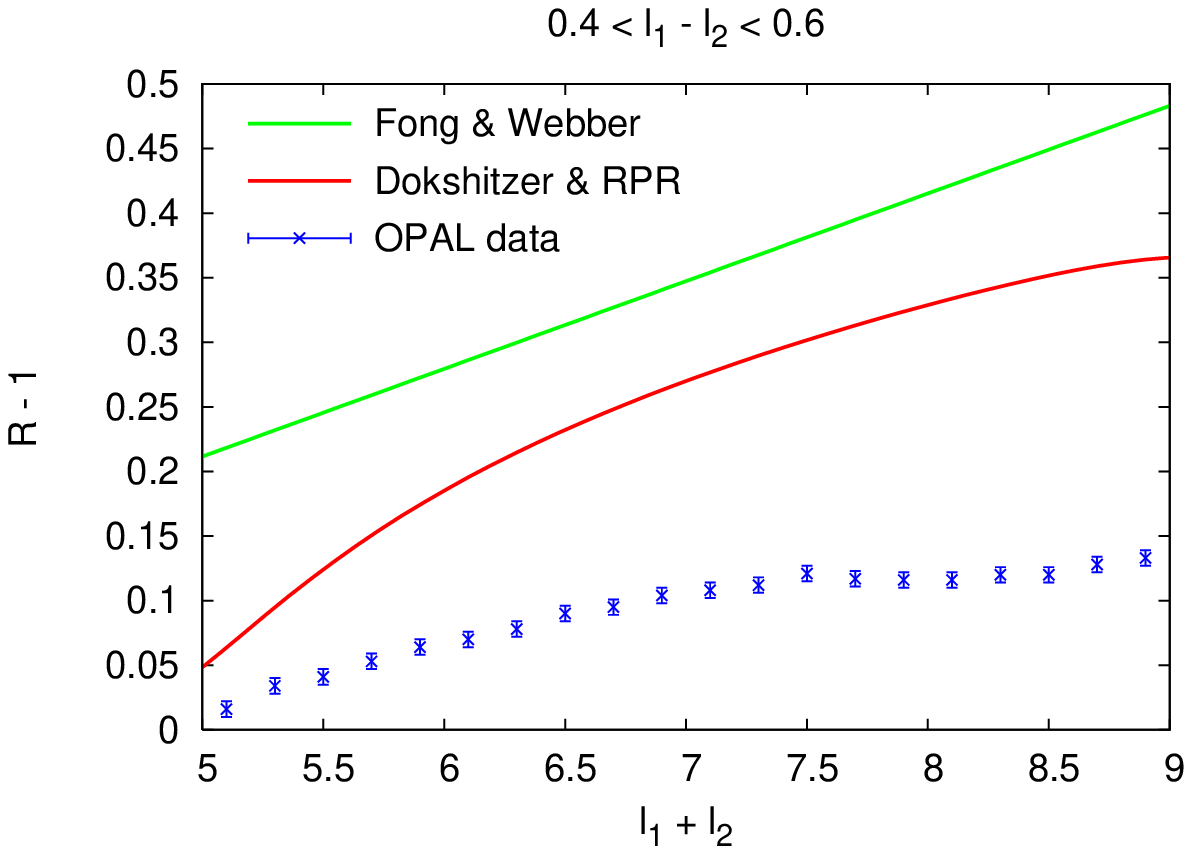, height=4truecm,width=0.45\tw}
\hfill
\epsfig{file=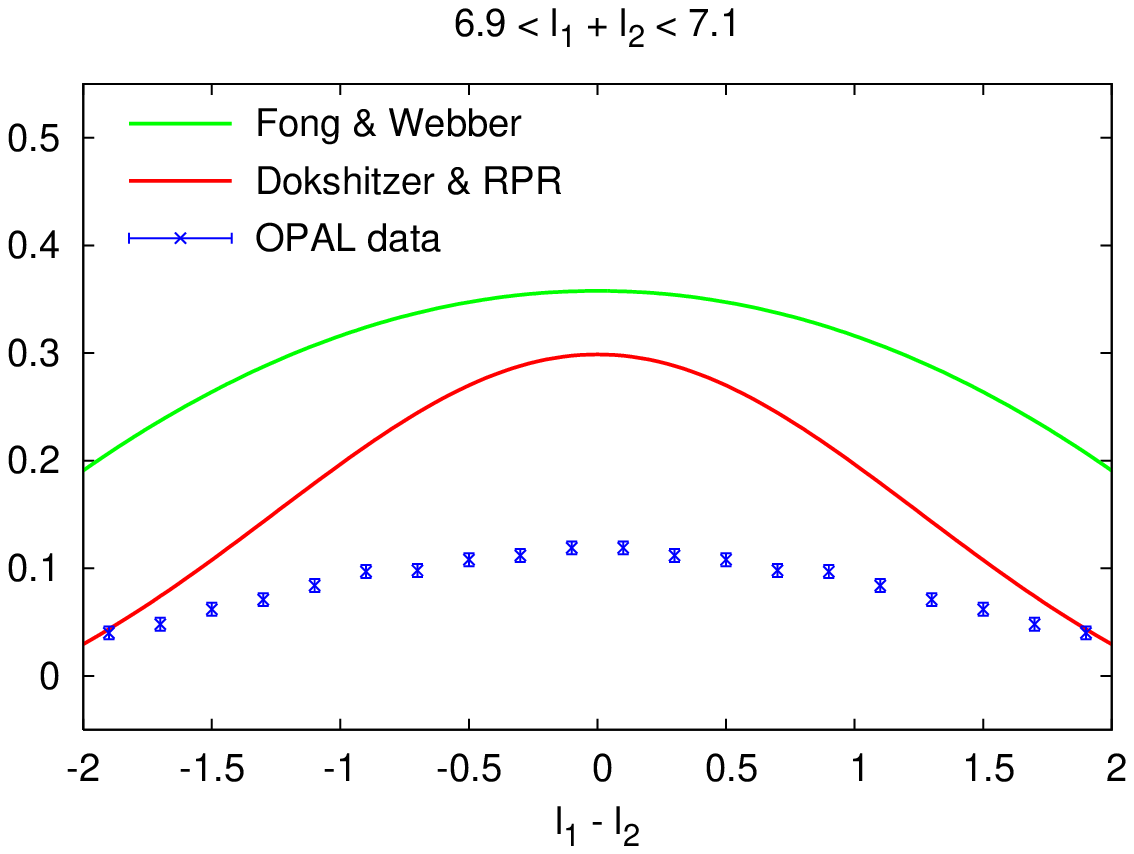, height=4truecm,width=0.45\tw}
\end{center}

\begin{center}
\epsfig{file=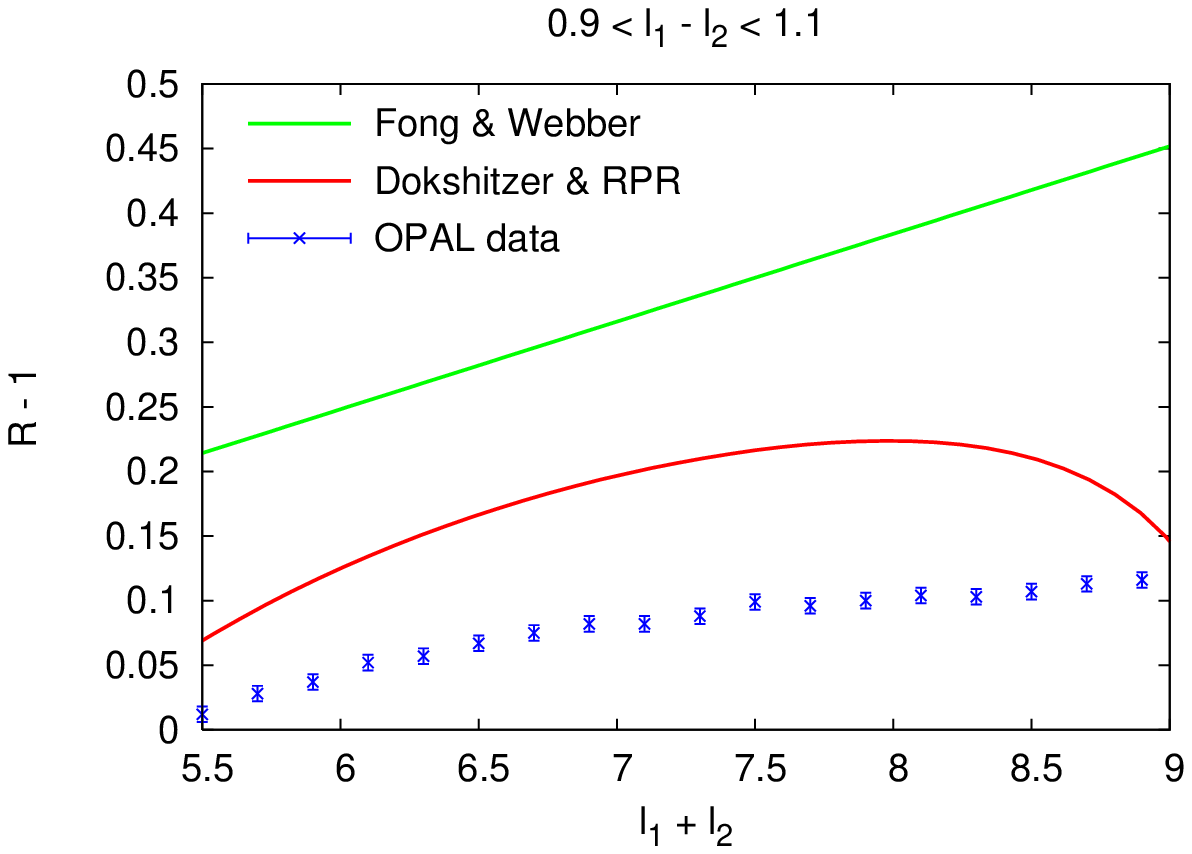, height=4truecm,width=0.45\tw}
\hfill
\epsfig{file=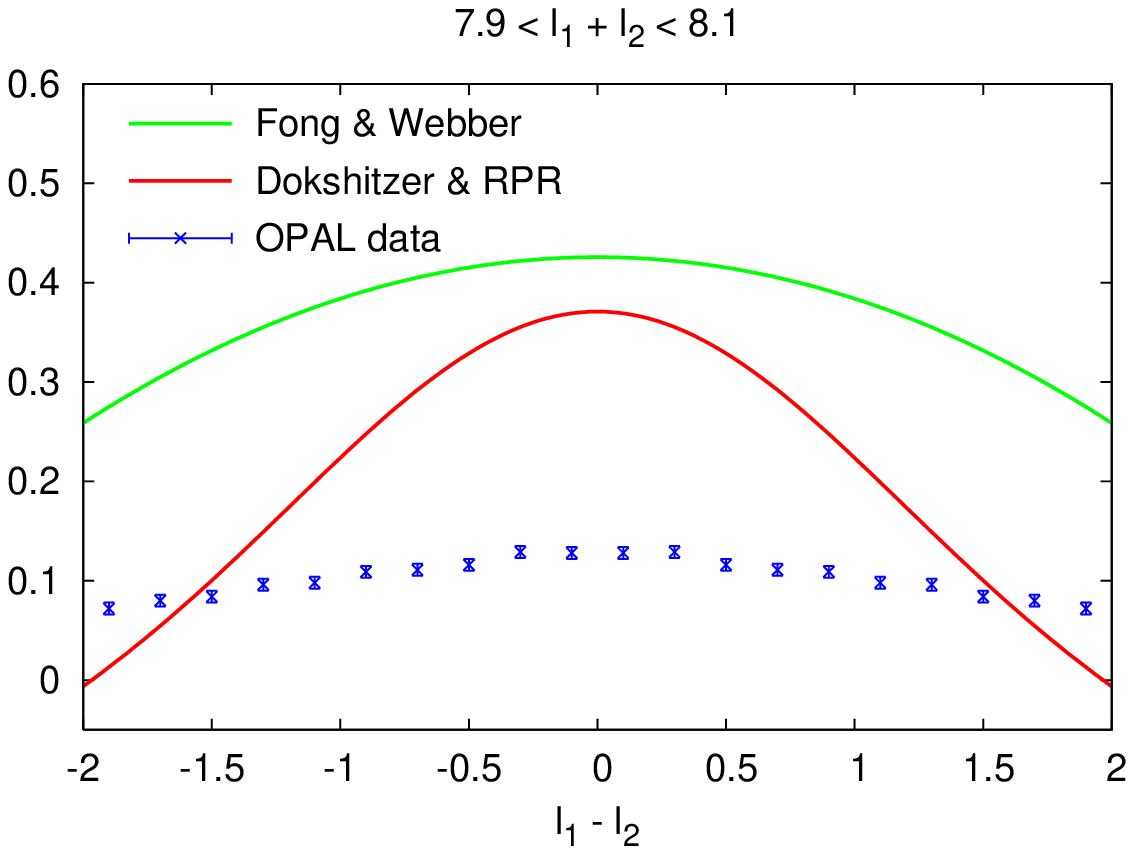, height=4truecm,width=0.45\tw}
\end{center}

\vspace {2mm}
\noindent
{\large\bf Conclusions}

The correlation should be strongest when the two particles have the
same energy, $\ell_1=\ell_2$, in agreement with the Fong--Webber
analysis.

At the same time, the correlation function that we derived directly
from the evolution equations without expanding the answer in the
formal perturbation parameter $\gamma_0$, no longer increases linearly
but flattens off (and then tends to decrease) with $\ell_1+\ell_2$.
Such a behaviour is in accord with general theoretical expectations.
Indeed, the smallest energy gluon with $\ell\to Y$ is pushed at large
angles, $\Theta\sim1$, and, in virtue of the QCD coherence, should be
radiated independently of the rest of the parton ensemble.

Though our curves are somewhat closer to the OPAL measurements than
the Fong--Webber results, the discrepancy remains substantial.
Whether this discrepancy is due to higher order perturbative
corrections or rather due to non-trivial hadronization effects that
have not been seen in the inclusive one-particle spectra, remains to
be studied.
Forthcoming experimental data on two-particle correlations in $g/q$
jets produced in $pp$ collisions (CDF) should elucidate this problem.


\begin{thebibliography}{9}
\bibitem{KO} V.A.\ Khoze and W.\ Ochs, Int.\ J.\ Mod.\ Phys. {\bf A12}
  (1997) 2949 .
 
\bibitem{DPR} 
Yu.L.\ Dokshitzer and R.\ Perez Ramos, under preparation

\bibitem{FW} C.P.\ Fong and B.R.\ Webber, Phys.\ Lett. {\bf B241} (1990) 255.

\bibitem{EvEq} 
Yu.L.\ Dokshitzer, V.A.\ Khoze, A.H.\ Mueller and S.I.\ Troyan,  \\
{\em Basics of Perturbative QCD}, Editions Frontieres, Paris, 1991.

\bibitem{DLA} Yu.L.\ Dokshitzer, V.S.\ Fadin and V.A.\ Khoze, Z.\ Phys.
  {\bf C18} (1983) 37. 

\bibitem{OPAL} OPAL Collab.,Physics Letters B 287 (1992) 401-412

\end{thebibliography}
\end{document}